\documentclass[twocolumn,twocolappendix,resetfootnote]{aastex701}

\usepackage[table]{xcolor}
\usepackage{xcolor}

\usepackage{amsmath,amssymb,amsfonts}%
\usepackage{amsthm}%
\usepackage{mathrsfs}%
\usepackage{CJK}
\newcommand{\pc}{{\rm pc}}
\newcommand{\K}{{\rm K}}

\newcommand{\cc}{{\rm cm}^{-3}}
\newcommand{\mum}{\mu{\rm m}}
\usepackage{multirow}

\shorttitle{LRD free-free radio emission}
\shortauthors{Tanaka, Inayoshi, et al.}

\begin{document}
% \linenumbers
\begin{CJK*}{UTF8}{gbsn}

\title{Free-Free Radio Emission from Little Red Dots as a Probe of Ionized Gas}

\author[0009-0003-4742-7060]{Takumi S. Tanaka}
\affiliation{Kavli Institute for the Physics and Mathematics of the Universe (WPI), The University of Tokyo Institutes for Advanced Study, The University of Tokyo, Kashiwa, Chiba 277-8583, Japan}
\affiliation{Department of Astronomy, Graduate School of Science, The University of Tokyo, 7-3-1 Hongo, Bunkyo-ku, Tokyo 113-0033, Japan}
\affiliation{Center for Data-Driven Discovery, Kavli IPMU (WPI), UTIAS, The University of Tokyo, Kashiwa, Chiba 277-8583, Japan}
\email[show]{takumi.tanaka@ipmu.jp}

\author[orcid=0000-0001-9840-4959]{Kohei Inayoshi}
\correspondingauthor{Kohei Inayoshi}
\email[]{inayoshi@pku.edu.cn}
\affiliation{Kavli Institute for Astronomy and Astrophysics, Peking University, Beijing 100871, China}
\email[]{inayoshi@pku.edu.cn}

\author[orcid=0009-0008-9591-7052,sname='Yan']{Zu Yan (晏祖)}
\affiliation{Kavli Institute for Astronomy and Astrophysics, Peking University, Beijing 100871, China}
\affiliation{Department of Astronomy, School of Physics, Peking University, Beijing 100871, China}
\email{yanzu@stu.pku.edu.cn}  

\author[0009-0004-4332-9225]{Tomokazu Kiyota}
\affiliation{Astronomical Science Program, Graduate Institute for Advanced Studies, SOKENDAI, 2-21-1 Osawa, Mitaka, Tokyo 181-8588, Japan}
\affiliation{National Astronomical Observatory of Japan, 2-21-1 Osawa, Mitaka, Tokyo, 181-8588, Japan}
\email{tomokazu.kiyota@grad.nao.ac.jp}

\author[0000-0002-6047-430X]{Yuichi Harikane}
\affiliation{Institute for Cosmic Ray Research, The University of Tokyo, 5-1-5 Kashiwanoha, Kashiwa, Chiba 277-8582, Japan}
\email{hari@icrr.u-tokyo.ac.jp}

\author[0000-0002-0000-6977]{John D. Silverman}
\affiliation{Kavli Institute for the Physics and Mathematics of the Universe (WPI), The University of Tokyo Institutes for Advanced Study, The University of Tokyo, Kashiwa, Chiba 277-8583, Japan}
\affiliation{Department of Astronomy, Graduate School of Science, The University of Tokyo, 7-3-1 Hongo, Bunkyo-ku, Tokyo 113-0033, Japan}
\affiliation{Center for Data-Driven Discovery, Kavli IPMU (WPI), UTIAS, The University of Tokyo, Kashiwa, Chiba 277-8583, Japan}
\email{john.silverman@ipmu.jp}
 
\begin{abstract}
Recent studies have hypothesized that little red dots (LRDs) are rapidly accreting black holes surrounded by dense gaseous environments.
Both neutral and ionized gas play a key role in explaining many of the puzzling spectral features of LRDs, including the emission line profiles in the optical-infrared bands, although the physical conditions required for these gas components remain poorly constrained. 
We propose that free-free emission provides a robust, independent probe of ionized gas in LRDs.
Free-free self-absorption produces a characteristic spectral turnover whose critical frequency and luminosity depend on the electron column density and the characteristic size of the ionized region, such that radio observations simultaneously constrain both quantities.
The spectral slope at frequencies below the turnover further probes the radial density distribution on spatial scales far below those achievable by current and future optical-infrared observatories.
We also present predicted free-free spectral energy distributions for two representative LRDs in the local and high-redshift universe and show that future radio-to-millimeter observations with \textit{ALMA} and \textit{ngVLA} can probe the range of ionized-gas parameters inferred from current observations.
Free-free emission therefore offers a unique way to probe the geometry and physical conditions of the ionized gas surrounding LRDs.
\end{abstract}


\section{Introduction} \label{sec:introduction}
\end{CJK*}
The discovery of active galactic nuclei (AGNs) at high redshift with \textit{JWST} has provided a long-awaited opportunity to uncover their formation and early evolution \citep[e.g.,][]{Onoue2023, Harikane2023_agn, Maiolino2023, Maiolino2023_GNz11, Taylor2025, Tanaka2025_z10}.
Among them, a newly identified population of high-redshift, broad-line emitters known as ``little red dots'' (LRDs) has been actively discussed as a key for probing the rapid growth of massive black holes (BHs) in the early Universe (e.g., \citealt{Labbe2023a, Furtak2024, Greene2024, Matthee2024, Akins2024}; see \citealt{InayoshiHo2025_review} for a review).

LRDs are considered likely AGNs, as suggested by their broad emission lines and compact morphologies. 
However, their distinctive V-shaped rest-UV-to-optical spectra, which rise sharply toward wavelengths longer than the Balmer limit \citep[e.g.,][]{Greene2024,Wang2024b, Kokorev2024_break} and their unusually weak X-ray, hot-dust, and radio emission \citep{Williams2024,Maiolino2024_chandra,Akins2024,Setton2025} challenge the standard picture of AGN structures and stimulate the discussion of their interpretations.
Moreover, their cosmic abundance rises from high redshifts, peaks at $z\sim 6$ at a level two orders of magnitude higher than the quasar abundance \citep[e.g.,][]{Niida2020,Matsuoka2023}, and declines toward lower redshifts \citep[e.g.,][]{Kocevski2024, Kokorev2024_census,Rinaldi2026_evol}.
This transient yet dominant population may therefore represent a major phase of the early BH growth \citep{Inayoshi2025}.

A growing body of work has proposed that LRDs are deeply embedded in dense gas with a high covering fraction \citep[e.g.,][]{InayoshiMaiolino2024, Naidu2025, Kido2025, deGraaff2025_spec_stat}. 
Radiative-transfer effects in dense (partially ionized) neutral gas can produce the red optical continua, strong Balmer breaks, prominent Balmer absorption, and emission observed in LRDs \citep[e.g.,][]{deGraaff2025,Naidu2025,Kido2025,Chang2026,Yan2026_break,Kiyota2026}, while also suppressing the emergent X-ray radiation. 
Dense ionized gas with an electron column density of $N_{\rm e}\sim10^{24}\,{\rm cm}^{-2}$ has been proposed to broaden the Balmer emission lines through electron scattering \citep[e.g.,][]{Rusakov2025,Matthee2026_gasdensity}.
These effects can substantially reshape the observed spectra and may bias inferences of intrinsic properties, including single-epoch virial BH masses (\citealt{Rusakov2025}, but see \citealt{Brazzini2025,Juodzbalis2025}). 
Despite the growing evidence for multi-phase gas components, the physical conditions and the geometry of LRDs remain poorly constrained.

In this Letter, we consider free-free emission from ionized gas surrounding LRD nuclei.
This paper assumes an ionized gas component that extends outside the hypothesized photospheric envelope (Section\,\ref{ssec:scales}).
The free-free emission spectrum is determined by a relatively small number of physical parameters of the ionized gas, independent of the radiative-transfer signatures in the optical and near-infrared spectra (Section\,\ref{ssec:model}).
Therefore, future deep radio observations will provide direct constraints on the properties of the ionized gas, inaccessible from imaging analysis alone (Section\,\ref{sec:discussion}).

Throughout this work, we assume a flat $\Lambda$CDM cosmology with $H_0 = 67.4\,{\rm km\,s^{-1}\,Mpc^{-1}}$, $\Omega_{\rm m} = 0.315$, and $\Omega_\Lambda=0.685$ \citep{Planck2018}.

\vspace{5mm}
\section{Free-free emission SEDs} \label{sec:model}

\subsection{Physical scales}\label{ssec:scales}

Radiation spectra of LRDs are often interpreted as a combination of quasi-thermal emission with $T_{\rm eff}\simeq 5000\,\K$ from a dense, partially ionized gaseous envelope.
The detection of [O\,{\sc iii}] and other metal emission lines indicates the presence of ionizing photons with energies above $h\nu_{\rm H}=13.6\,{\rm eV}$ produced in the central engine.
For a typical UV luminosity of $M_{\rm UV}\simeq -18\,{\rm mag}$, the ionizing photon production rate is estimated to be $Q_{\rm H}\simeq 4\times10^{53}\,{\rm s}^{-1}$, assuming an ionizing photon production efficiency of $\log\xi_{\rm ion}\simeq 25.8$, comparable to the high values inferred for blue, strong-line-emitting galaxies \citep[e.g.,][]{Bouwens2016}.

If the LRD is embedded in a gaseous medium with a uniform hydrogen number density $n_{\rm H}$, the radius of the surrounding H\,{\sc ii} region is given by
\begin{align}
    R_{\rm ion}&=\left(\frac{3Q_{\rm H}}{4\pi n_{\rm H}^2 \alpha_{\rm B}}\right)^{1/3},\nonumber \\
   &\simeq 3.2\,\pc\,Q_{\rm H,53}^{1/3} \, n_{\rm H,4}^{-2/3} \, T_4^{0.28}, \label{eq:scale_R}
\end{align}
where $Q_{\rm H,53}=Q_{\rm H}/(10^{53}\,{\rm s}^{-1})$, $n_{\rm H,4}=n_{\rm H}/(10^4\,\cc)$, and $T_4=T/(10^4\,\K)$.
The corresponding electron column density is
\begin{align}
    N_{\rm e}\simeq 0.97\times 10^{23}\,{\rm cm}^{-2}
   \,Q_{\rm H,53}^{1/3} \, n_{\rm H,4}^{1/3} \, T_4^{0.28}. \label{eq:scale_N}
\end{align}
Here, we primarily consider an ionized gas component that can also produce the observed [O\,{\sc iii}]\,$\lambda5007$ emission.
We therefore assume an electron density below the critical density for collisional de-excitation of this transition, $\lesssim 7\times10^5\,{\rm cm}^{-3}$ \citep{Baskin2005}.
The corresponding electron-scattering optical depth across this pc-scale ionized region is of order $\tau_{\rm es}\sim\mathcal{O}(0.1-1)$, broadly consistent with or somewhat smaller than the optical depths inferred from the exponential wings of broad Balmer emission lines when electron scattering is the dominant broadening mechanism.
Note that the ionized gas considered here need not be identical to the much more compact, denser gas responsible for the broad Balmer-line wings through electron scattering \citep{Rusakov2025,Sneppen2026}.

\subsection{Spectral models}\label{ssec:model}

We calculate the free-free emission from the LRD nuclei.
Free-free emission from compact H\,{\sc ii} regions has long been recognized as a useful probe of the plasma density, temperature, and geometry \citep[e.g.,][]{Olnon1975}.
We assume an axisymmetric plasma distribution along the line of sight (LoS) with uniform electron temperature and a prescribed profile.
Let $z$ denote the depth along the LoS measured from the central radiation source and let $\varpi$ be the cylindrical radius perpendicular to the LoS. The spherical radius is then given by $r=\sqrt{\varpi^2+z^2}$.
The dominant opacity source is free-free absorption (${\rm H}^++{\rm e}^-+\gamma \rightarrow {\rm H}^++{\rm e}^-+\gamma'$) and the optical depth along a ray with impact parameter $\varpi$ is 
\begin{equation}
    \tau_\nu(\varpi) = 2f(\nu, T_{\rm e})\int_0^\infty n_{\rm e}^2(r)dz,
\end{equation}
where $f(\nu,T_{\rm e})$ contains the frequency and temperature dependence of the free-free absorption coefficient, including the Gaunt factor.
Throughout this paper, we adopt the approximation of \cite{Altenhoff1960} and \cite{Mezger1967},
\begin{equation}
    f(\nu, T_{\rm e})=3.28\times 10^{-11}\,{\rm cm}^6\,\pc^{-1}
    \,T_{\rm e,4}^{-1.35}
    \left(\frac{\nu}{100\,{\rm GHz}}\right)^{-2.1}.
    \label{eq:FF}
\end{equation}

In this paper, we consider two models for the plasma density distribution.
First, we assume a uniform-density sphere (Model II of \citealt{Olnon1975}), in which $n_{\rm e}=n_0$ inside a sphere of radius $R$ and $n_{\rm e}=0$ outside.
The luminosity density is
\begin{align}
L_\nu &= \frac{8\pi^2 k_{\rm B}T_{\rm e}R^2\nu^2}{c^2}\left[1-\frac{2}{p_\nu^2}\left\{1-(p_\nu+1)e^{-p_\nu}\right\}\right], \label{eq:model_ii} \\[5pt]
&=\frac{8\pi^2 k_{\rm B}T_{\rm e}R^2}{c^2}\times 
\begin{cases}
    \frac{4}{3}n_0^2R \nu^2 f & \text{for $\nu\gg \nu_{\rm c}$,} \\
    \nu^2                    & \text{for $\nu\ll \nu_{\rm c}$,} \\
  \end{cases}\nonumber
\end{align}
where $p_\nu = 2n_0^2Rf$.
The two asymptotic limits correspond to the optically thin ($\propto \nu^{-0.1}$) and thick ($\propto \nu^2$) regimes, respectively.
The critical frequency is defined by $p(\nu_{\rm c})=1$.

As an alternative model, we consider a truncated power-law density profile of $n_{\rm e}(r)=n_0(r/R)^{-q}$ at $r\geq R$ with a homogeneous sphere with $n_0$ at $r<R$ (Model V of \citealt{Olnon1975}). 
For the representative case of $q=2$, the luminosity density is calculated as
\begin{align}
L_\nu =\frac{8\pi^2 k_{\rm B}T_{\rm e}R^2}{c^2}\times 
\begin{cases}
    \frac{16}{3}n_0^2R \nu^2 f & \text{for $\nu\gg \nu_{\rm c}$,} \\
    \nu^2\left(\frac{\pi}{2}n_0^2Rf\right)^{2/3}\Gamma(\frac{1}{3}) & \text{for $\nu\ll \nu_{\rm c}$,} \\
  \end{cases}
  \label{eq:model_v}
\end{align}
where $\Gamma$ is the Gamma function.
As in the uniform-density model, the plasma becomes optically thin at sufficiently high frequencies, yielding $F_\nu\propto\nu^{-0.1}$. 
At low frequencies, however, the density gradient modifies the emergent spectrum to $F_\nu\propto\nu^{0.6}$, which is substantially shallower than the optically thick spectrum of a uniform sphere.
More generally, the low-frequency spectral index is $\alpha = (2q-3.1)/(q-0.5)$,
which approaches the uniform-density case, $\alpha\rightarrow2$, as $q\rightarrow\infty$.

\subsection{Physical constraints from free-free emission}\label{ssec:implication}

\begin{figure}
\centering
\includegraphics[width=0.98\linewidth]{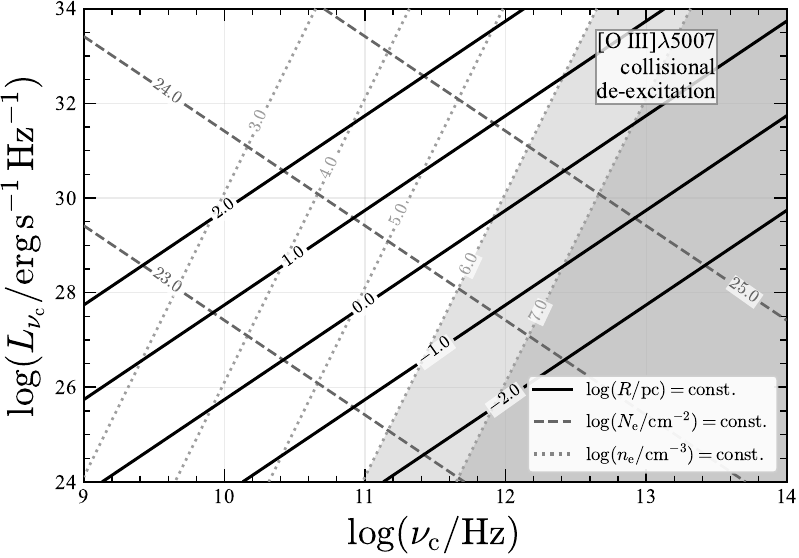}
\caption{
The relation between the free-free emission critical frequency, $\nu_{\rm c}$, and the luminosity at $\nu_{\rm c}$, $L_{\nu_{\rm c}}$.
Solid, dashed, and dotted contours indicate constant ionized-region radius, $\log\left(R/{\rm pc}\right)$, electron column density, $\log\left(N_{\rm e}/{\rm cm^{-2}}\right)$, and electron number density, $\log\left(n_{\rm e}/{\rm cm^{-3}}\right)$, respectively.
The gray shaded regions indicate $\log\left(n_{\rm e}/{\rm cm^{-3}}\right)>6$ and $>7$, respectively, where collisional de-excitation of [O\,{\sc iii}]\,$\lambda5007$ becomes non-negligible.
We assume a uniform-density ionized region (Model II, where $N_{\rm e}=n_{\rm e} R$) with $T_{\rm e}=10^4\,{\rm K}$.
Measurements of $\nu_{\rm c}$ and $L_{\nu_{\rm c}}$ can constrain both $N_{\rm e}$ and $R$.
}
\label{fig:Lnuc_nuc}
\end{figure}

For the uniform-density case, the critical frequency is approximately evaluated as 
\begin{equation}
\nu_{\rm c}\simeq 26\,{\rm GHz}\,T_{\rm e,4}^{-3/4} \, N_{\rm e,23} \, R_{\rm 1\,pc}^{-1/2},
\label{eq:nuc}
\end{equation}
where $N_{\rm e,23}=N_{\rm e}/(10^{23}\,{\rm cm}^{-2})$ and $R_{\rm 1\,pc}=R/(1\,{\rm pc})$.
The luminosity density at the critical frequency is roughly estimated as
\begin{align}
\label{eq:Lnuc}
L_{\nu_{\rm c}}  & =\frac{16\pi^2k_{\rm B}T_{\rm e}R^2}{3c^2}\nu_{\rm c}^2,\\
&\simeq 5.33\times 10^{26}\,{\rm erg\,s}^{-1}\,{\rm Hz}^{-1} \, T_{\rm e,4}^{-1/2} \, N_{\rm e,23}^{2} \,R_{\rm 1\,pc}.\nonumber
\end{align}
Equations\,(\ref{eq:nuc}) and (\ref{eq:Lnuc}) are derived under the assumption that the free-free absorption coefficient is approximated as $f(\nu,T_{\rm e})\propto T_{\rm e}^{-3/2}\nu^{-2}$, neglecting the $\nu$ and $T_{\rm e}$ dependence of the Gaunt factor\footnote{Only the power-law indices are modified, while the normalization in Equation\,(\ref{eq:FF}) is unchanged. The resulting difference is at the $\sim 1-2\%$ level.}.
When the electron temperature is relatively constrained, both the critical frequency and the corresponding luminosity can thus be expressed as $\nu_{\rm c}\propto N_{\rm e}R^{-1/2}$ and $L_{\nu_{\rm c}}\propto N_{\rm e}^2 R$, implying $N_{\rm e}\propto \nu_{\rm c}^{1/2} L_{\nu_{\rm c}}^{1/4}$ and $R\propto \nu_{\rm c}^{-1} L_{\nu_{\rm c}}^{1/2}$.
Figure\,\ref{fig:Lnuc_nuc} illustrates the mapping between the observable quantities for the free-free emission, $(\nu_{\rm c},L_{\nu_{\rm c}})$, and the physical properties of the ionized gas, $(N_{\rm e},R)$.
Constraints on $\nu_{\rm c}$ and $L_{\nu_{\rm c}}$ therefore provide estimates of both the electron column density and the characteristic size of the ionized region.
In addition, the spectral slope on the low-frequency side at $\nu<\nu_{\rm c}$ provides an independent constraint on the radial density profile of the ionized gas on spatial scales far below the spatial resolution of current and future observatories.

Even if we treat $T_{\rm e}$ as a free parameter, its impact on the inferred $N_{\rm e}$ and $R$ remains relatively small. 
As shown in Equations\,(\ref{eq:nuc}) and (\ref{eq:Lnuc}), the degeneracies with the electron temperature follow $N_{\rm e}\propto T_{\rm e}^{1/2}$ and $R\propto T_{\rm e}^{-1/2}$.
Here, we assume a conservative parameter range of $5\times10^{3}<T_{\rm e}/{\rm K}<3\times10^{4}$, which encompasses a broad range of temperatures of photoionized gas in both star-forming H\,{\sc ii} regions and AGN narrow-line regions \citep[e.g.,][]{Osterbrock2006, Zhang2013_SDSS}.
The lower bound is of the same order as the typical $T_{\rm eff}$ inferred for the dense, partially ionized gas envelopes of LRDs \citep[e.g.,][]{InayoshiMaiolino2024,deGraaff2025_spec_stat,Umeda2025_BHstar,Perez-Gonzalez2026_stack}, while the upper bound is consistent with recent direct-$T_{\rm e}$ measurements of narrow-line gas in LRDs \citep{Nikopoulos2026}.
This $T_{\rm e}$ range corresponds to a variation of approximately $\pm 0.4\,{\rm dex}$ in $T_{\rm e}$, leading to an uncertainty of approximately $\pm 0.2\,{\rm dex}$ in the inferred $N_{\rm e}$ and $R$.
This is comparable to the $N_{\rm e}$ constraints from optical line profile analyses by \cite{Rusakov2025}, which typically have uncertainties of $\pm 0.1\,{\rm dex}$.

We note that throughout the discussion above, we have focused on the Rayleigh-Jeans regime in order to simplify the equations in free-free emission and absorption coefficients.
At sufficiently high frequencies, the spectrum exhibits an exponential cutoff due to the thermal electron distribution, and the free-free emission scales as
$F_\nu \propto \nu^{-0.1}e^{-h\nu/(k_{\rm B}T_{\rm e})}$.
This suppression becomes important at frequencies above $\nu_{\rm m}\equiv k_{\rm B}T_{\rm e}/h=2.1\times 10^{14}\,{\rm Hz}$, corresponding to a wavelength of $\lambda_{\rm m}\sim 1.4\,\mum$.
Therefore, free-free emission from the ionized region yields an approximately flat spectrum in $F_\nu$ at rest-frame wavelengths $\lambda\gtrsim \lambda_{\rm m}$, followed by an exponential decline at higher frequencies.
This flat spectrum in rest-frame near-infrared bands produces an unavoidable contribution of thermal bremsstrahlung in ionized gas surrounding LRD nuclei (see also \citealt{Naidu2026_Carina}).

\section{Discussion} \label{sec:discussion}

\begin{figure*}
\centering
\includegraphics[width=0.95\linewidth]{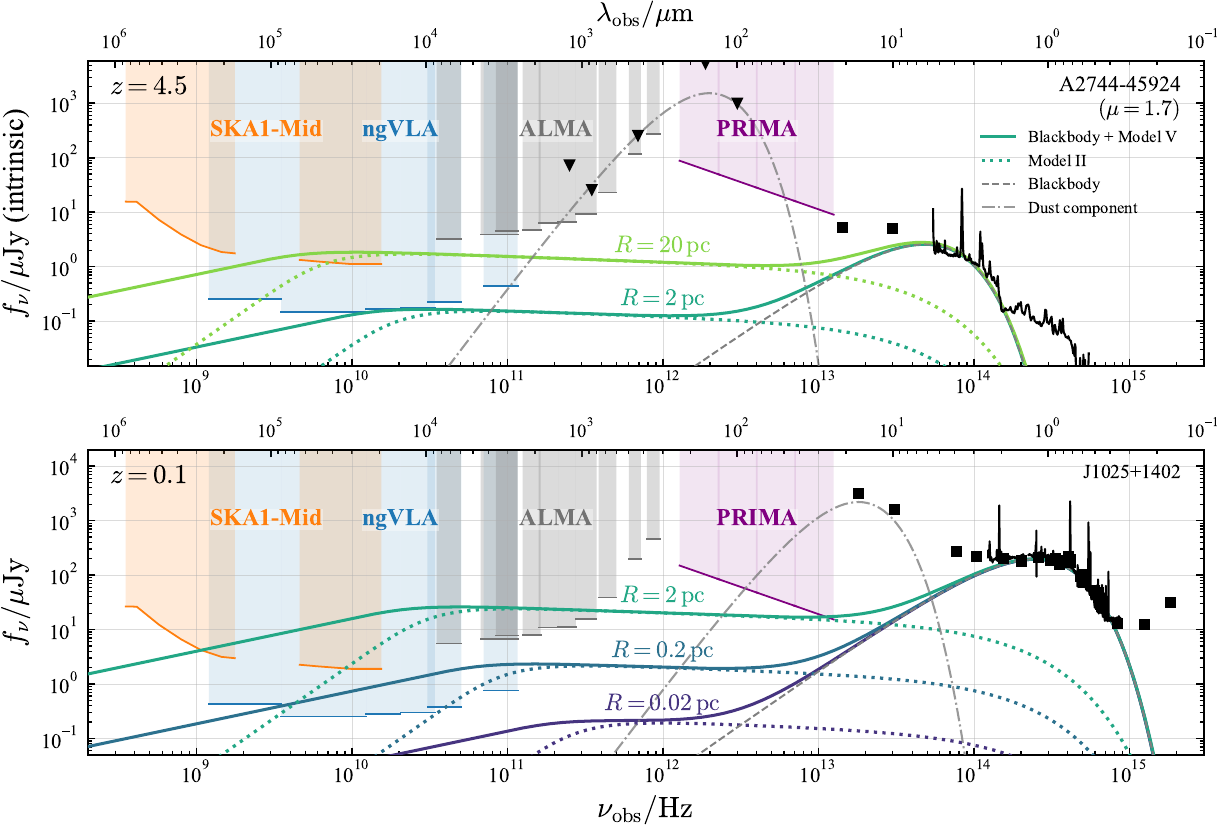}
\caption{
Expected free-free SEDs for representative high- and low-redshift LRDs.
The upper panel shows the $z=4.5$ case, calculated with $T_{\rm e}=1\times10^4\,{\rm K}$, $N_{\rm e}=1\times10^{24}\,{\rm cm^{-2}}$, and $R=2$ and $20\,{\rm pc}$.
The lower panel shows the $z=0.1$ case, calculated with $T_{\rm e}=1\times10^4\,{\rm K}$, $N_{\rm e}=3\times10^{23}\,{\rm cm^{-2}}$, and $R=0.02$, $0.2$, and $2\,{\rm pc}$.
The solid curves show the sum of the Model V free-free component and an illustrative blackbody component for reproducing the rest-optical component, while the dotted curves show the Model II free-free component.
We overplot the observed spectra and photometry of A2744-45924 \citep{Setton2025} at $z=4.46$ and J1025+1402 \citep{Lin2025_lowz} at $z=0.1007$.
The observed flux densities of A2744-45924 are corrected for a lensing magnification of $\mu=1.7$.
The horizontal lines and shaded regions indicate the $5\sigma$ continuum sensitivities expected from 10-hour integrations with \textit{SKA1-Mid}, \textit{ngVLA}, \textit{ALMA}, and \textit{PRIMA}.
The dashed and dash-dotted curves show blackbody and modified-blackbody components with $\beta=1.8$ and $T_{\rm dust}=110$ and $200\,{\rm K}$ for the $z=4.5$ and $z=0.1$ cases, respectively, consistent with the current SED constraints.
}
\label{fig:SED}
\end{figure*}

\begin{figure*}
\centering
\includegraphics[width=0.85\linewidth]{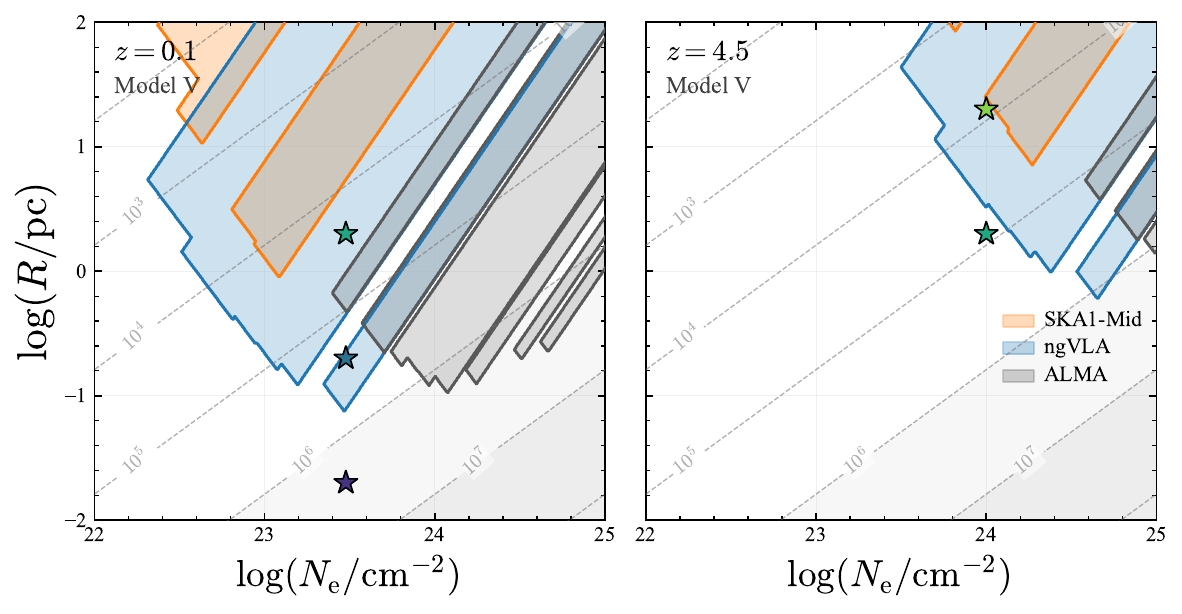}
\caption{
The $(N_{\rm e},R)$ parameter space in which $L_{\nu_{\rm c}}$ can be detected at the $5\sigma$ level with 10-hour exposures using \textit{SKA1-Mid}, \textit{ngVLA}, and \textit{ALMA}.
The left and right panels correspond to $z=0.1$ and $z=4.5$, respectively.
The stars indicate the assumed parameters in the example SEDs shown in Figure\,\ref{fig:SED}. 
}
\label{fig:model_detectability}
\vspace{5mm}
\end{figure*}

\begin{figure}
\centering
\includegraphics[width=0.95\linewidth]{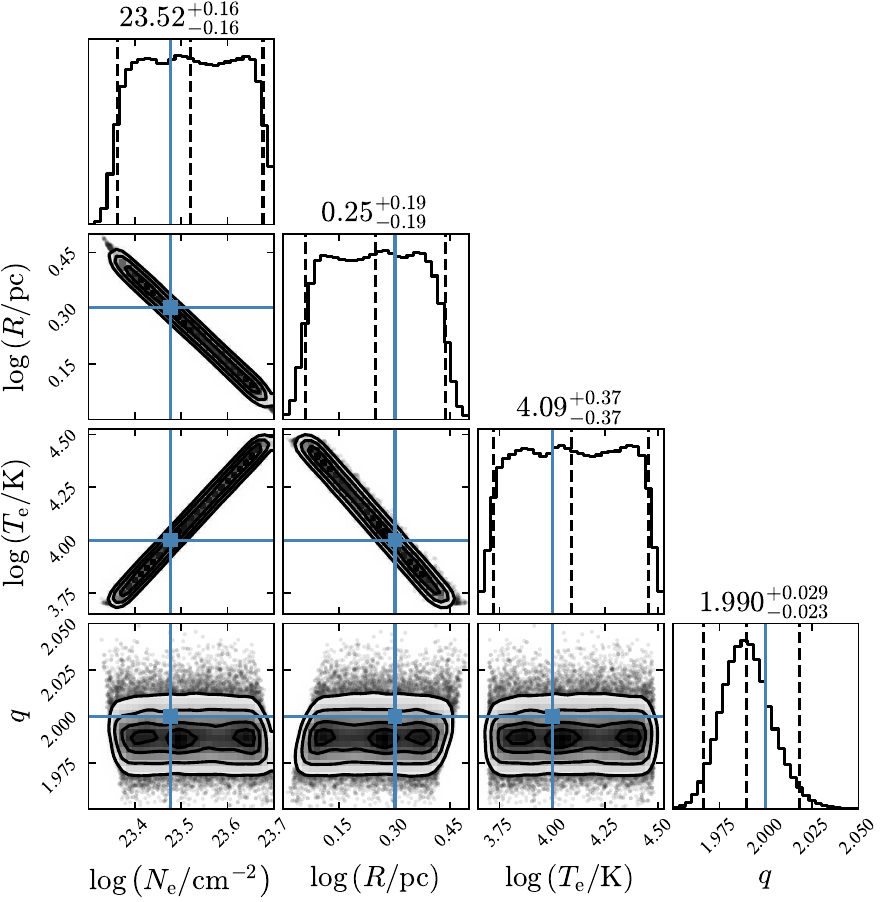}
\caption{
Posterior distributions obtained by fitting the mock photometry with the Model V free-free emission model.
Blue dots and solid lines indicate the true values assumed in the mock data generation.
Black dashed lines and the top labels indicate the median and the $2\sigma$ confidence intervals.
We cannot solve the degeneracy among $T_{\rm e}$, $N_{\rm e}$, and $R$ in principle.
However, $N_{\rm e}$ and $R$ can still be constrained to a precision of approximately $\pm0.2\,{\rm dex}$ because the physically plausible range of $T_{\rm e}$ is limited and $N_{\rm e}$ and $R$ depend only weakly on $T_{\rm e}$.
}
\label{fig:MCMC}
\end{figure}

\subsection{Detectability with future observations}

\subsubsection{Parameter settings and sensitivities}
In this Letter, we present high- and low-redshift case studies at $z=4.5$ and $0.1$, respectively, to investigate the expected free-free SEDs of LRDs and the physical constraints that can be obtained from future observations.
For the high-redshift case, we consider A2744-45924 \citep[e.g.,][]{Greene2024, Labbe2024}, which is one of the most luminous and well-studied LRDs at $z=4.46$ with extensive multi-wavelength coverage as a single object \citep{Setton2025}. 
We consider J1025+1402, a local LRD analog known as ``The Egg'' at $z=0.1007$ \citep[e.g.,][]{Lin2025_lowz, Ji2026_lord}.

For a given density profile, $N_{\rm e}$, $R$, and $n_0$ are related to one another.
In addition, the characteristic size of the ionized region can be approximately estimated from the ionizing photon rate $Q_{\rm H}$ and the gas density for a uniform medium, giving $R\approx R_{\rm ion}$ as an order-of-magnitude estimate.
However, more accurate estimates can depend sensitively on the density distribution, dust absorption of ionizing photons, and the geometry of the surrounding medium.
Rather than imposing the specific values of $R_{\rm ion}$ and $N_{\rm e}$ predicted by Equations~(\ref{eq:scale_R}) and (\ref{eq:scale_N}), we therefore treat $N_{\rm e}$ and $R$ as model parameters and explore a broad range around these expected values.
For simplicity, we fix the electron temperature to $T_{\rm e}=10^4\,{\rm K}$, a representative value for ionized gas in H\,{\sc ii} regions, in the following analysis.

% Note that the physical scales of the ionized gas discussed in Section\,\ref{ssec:scales} can depend sensitively on the gas density distribution, dust absorption of ionizing photons, and the geometry of the surrounding medium.
% We therefore do not restrict the models to $R_{\rm ion}$ and $N_{\rm e}$ predicted by Equations\,(\ref{eq:scale_R}) and (\ref{eq:scale_N}), but instead explore a broad range of $N_{\rm e}$ and $R$.
% Observationally, the size of the ionized region is constrained only indirectly, with the morphology of LRDs providing an approximate upper limit of $R\lesssim100\,{\rm pc}$ \citep[e.g.,][]{Labbe2023b,Furtak2023}.
% \fi

Sufficiently high $N_{\rm e}$ models with small $R$ imply very high electron densities $n_{\rm e}$.
As shown in Figure\,\ref{fig:Lnuc_nuc}, for $\log\left(N_{\rm e}/{\rm cm}^{-2}\right)\gtrsim23.5$, models with $\log\left(R/{\rm pc}\right)=-1$ and $-2$ correspond to $\log\left(n_{\rm e}/{\rm cm^{-3}}\right) \gtrsim 6$ and $7$, respectively.
At such high densities, collisional de-excitation of forbidden lines such as [O\,{\sc iii}]\,$\lambda5007$ may no longer be negligible \citep[e.g.,][]{Baskin2005, Maiolino2026}.
Thus, if the observed [O\,{\sc iii}]\,$\lambda5007$ emission originates from the ionized gas component considered here, such compact, high-density configurations are disfavored.
The observed [O\,{\sc iii}] emission could instead arise from more diffuse gas on larger, unresolved scales, which would not necessarily contribute significantly to the free-free emission or column density considered here.
Observationally, the spatial extent of the LRD is constrained only indirectly, with its unresolved morphology in the JWST/F444W imaging providing an approximate upper limit of $R\lesssim100~{\rm pc}$ \citep[e.g.,][]{Labbe2023b,Furtak2023}.
Therefore, although the [O\,{\sc iii}] emission provides a useful physical consideration for the compact ionized component, it does not by itself impose a stringent constraint on $R$.

To discuss the detectability of expected free-free emission, we consider the expected sensitivities of \textit{SKA1-Mid} (the first phase of Square Kilometre Array-Mid; \citealt{Braun2019}), \textit{ALMA}, and the \textit{ngVLA} (next generation Very Large Array; \citealt{ngvla_2018}).
\textit{SKA1} is currently under construction, and \textit{ngVLA} is planned to begin scientific operations in the 2030s.
For \textit{SKA1-Mid}, we estimate the 10-hour continuum sensitivities based on the expected instrumental performance reported by \cite{Braun2019}.
The sensitivities for \textit{ALMA} and \textit{ngVLA} are calculated using the ALMA Sensitivity Calculator\footnote{\url{https://almascience.org/proposing/sensitivity-calculator}} and the ngVLA exposure calculator\footnote{\url{https://ngect.nrao.edu/}}, respectively.

\subsubsection{Comparison results}

In Figure\,\ref{fig:SED}, we present the free-free emission spectra for the two LRDs described above: A2744-45924 at $z\simeq 4.5$ (top panel) and J1025+1402 at $z\simeq 0.1$ (bottom panel), together with the 5$\sigma$ continuum sensitivities expected from 10-hour on-source integrations in each band of ALMA and ngVLA.
The SEDs in both cases exhibit a characteristic turnover at $\nu\sim\nu_{\rm c}$. 
At $\nu>\nu_{\rm c}$, the optically thin emission follows a nearly flat spectrum of $F_\nu\propto\nu^{-0.1}$, whereas at $\nu<\nu_{\rm c}$, the spectral slope depends on the assumed radial density profile (see the discussions in Section\,\ref{sec:model}).

For A2744-45924, the free-free spectrum around $\nu\simeq\nu_{\rm c}$ is detectable with the \textit{ngVLA} for $R \gtrsim 2\,{\rm pc}$.
For J1025+1402, combining \textit{ngVLA} and \textit{ALMA} observations can detect the turnover with $R \gtrsim 1\,{\rm pc}$.
Even with the \textit{ngVLA} alone, the low-frequency part ($\nu<\nu_{\rm c}$) of the Model\,V SED remains detectable for $R \gtrsim 0.1\,{\rm pc}$.
These size constraints are much smaller than the expected spatial resolution of \textit{ngVLA}, which is $\sim 10^{-3}\,{\rm arcsec}$ at $\sim10^{2}\,{\rm GHz}$, corresponding to $\sim2\,{\rm pc}$ and $\sim7\,{\rm pc}$ at $z=0.1$ and $z=4.5$, respectively.

Figure\,\ref{fig:model_detectability} shows the $N_{\rm e}\,\mathchar`-\,R$ parameter space in which the free-free luminosity at the critical frequency, $L_{\nu_{\rm c}}$, can be detected at the $5\sigma$ level with 10-hour on-source integrations in the relevant bands of \textit{SKA1-Mid}, \textit{ALMA}, and \textit{ngVLA}.
In particular, at $z=0.1$, the \textit{ngVLA} can detect $L_{\nu_{\rm c}}$ over a broad range of the parameter space, highlighting the strong potential of local LRD analogs for directly constraining the properties of ionized gas.

To quantitatively assess the constraints that could be obtained from future observations, we fit mock photometry generated from a free-free SED.
We assume a Model V spectrum with $T_{\rm e}=1\times10^{4}\,{\rm K}$, $N_{\rm e}=3\times10^{23}\,{\rm cm^{-2}}$, $R=2\,{\rm pc}$, and $q=2$ at $z=0.1$, and generate mock photometry for 10-hour integrations in \textit{ngVLA} Bands-$2$, $4$, and $6$ and \textit{ALMA} Bands-$4$,  $6$, $7$, and $8$.
We include only free-free emission in the mock photometry; possible contamination from other emission components is discussed in Section\,\ref{ssec:contami}.
We then fit the mock photometry with a Model V SED with four free parameters, $\log\left(T_{\rm e}/{\rm K}\right)$, $\log\left(N_{\rm e}/{\rm cm^{-2}}\right)$, $\log\left(R/{\rm pc}\right)$, and $q$, using Markov chain Monte Carlo (MCMC) methods with \texttt{emcee} \citep{emcee}.
We adopt log-uniform priors of $\left[5\times10^3,\,3\times10^4\,{\rm K}\right]$ for $T_{\rm e}$, $\left[1\times10^{22},\,1\times10^{26}\,{\rm cm^{-2}}\right]$ for $N_{\rm e}$, and $\left[1\times10^{-3},\,1\times10^{2}\,{\rm pc}\right]$ for $R$, respectively.

The resulting posterior distributions are shown in Figure\,\ref{fig:MCMC}.
As discussed in Section\,\ref{ssec:implication}, the uncertainties in $N_{\rm e}$ and $R$ remain at approximately $\pm0.2\,{\rm dex}$ despite their degeneracy with $T_{\rm e}$.
Thus, observations with \textit{ALMA} and \textit{ngVLA} could provide an independent observational test for the presence of dense ionized gas and constrain its column density and size.

\subsection{Possible contamination from other emission}\label{ssec:contami}

\subsubsection{Thermal dust emission}
One possible contaminating component is thermal emission from cold dust.
For a pure blackbody in dust, the Rayleigh-Jeans tail follows $F_\nu\propto\nu^2$, which is comparable to the low-frequency slope of the free-free emission from Model\,II (a top-hat uniform density profile) at $\nu<\nu_{\rm c}$.
In contrast, a modified blackbody commonly adopted to describe dust emission has a steeper spectral slope, $F_\nu\propto\nu^{2+\beta}$ ($\beta>0$).
For Model\,V (a top-hat profile plus a power-law density tail), a more gradual decline in density outside the core results in a shallower free-free spectral slope at $\nu<\nu_{\rm c}$, making it more distinguishable from cold-dust emission.

For A2744-45924, the existing \textit{ALMA} upper limit places a strong constraint on the cold-dust emission.
In Figure\,\ref{fig:SED}, we add a modified-blackbody component with fiducial parameters of $T_{\rm dust}=110\,{\rm K}$, $\beta=1.8$, and $L_{\rm dust}=2\times10^{12}L_\odot$, chosen such that the model remains consistent with the \textit{ALMA} and \textit{Herschel} upper limits \citep{Setton2025}.
With these parameters, the dust contribution does not significantly obscure the free-free turnover for models with $R\gtrsim2\,{\rm pc}$.

For J1025+1402, we add a modified-blackbody component with $T_{\rm dust}=200\,{\rm K}$, $\beta=1.8$, and $L_{\rm dust}=3\times10^{9}L_\odot$, chosen such that the model is consistent with the existing \textit{WISE} photometry \citep{Lin2025_lowz}.
However, the cold-dust component is not well constrained due to the lack of mid-to-far infrared observations. 
Future \textit{ALMA} observations (e.g., ALMA Cycle 13 2026.1.00880.S; PI T.~Kiyota) could constrain both the cold-dust continuum and the free-free emission, thereby clarifying the detectability of analogous free-free signatures with the \textit{ngVLA} in high-redshift LRDs.

In the longer term, \textit{PRIMA} (Probing Far-Infrared Mission for Astrophysics, see \citealt{PRIMA_GO2}), which is planned for the 2030s, would cover the mid-to-far-infrared wavelength range of $24\,\mathchar`-\,235\,{\rm \mu m}$ with much higher sensitivity than \textit{Herschel}.
We estimate the continuum sensitivity for a 10-hour \textit{PRIMA}/FIRESS low-resolution point-source observation using the PRIMA Exposure Time Calculator\footnote{\url{https://prima.ipac.caltech.edu/page/etc-calc}}, assuming the $R=10$ binning.
As indicated by the purple lines in Figure\,\ref{fig:SED}, \textit{PRIMA} would improve constraints on the LRD dust continuum even at $z\sim5$, enabling a clearer separation of the cold-dust and free-free emission components.

\subsubsection{Synchrotron emission}
Another possible contamination component is synchrotron emission associated with star formation or an AGN jet.
Optically thin synchrotron emission generally exhibits a negative spectral slope with $F_\nu\propto\nu^{-0.7}$ \citep{Condon1992}.
Indeed, \cite{Rodriguez2026} reported a \textit{VLA} detection for another local LRD at $z=0.1682$ over $\simeq 4\,\mathchar`-\,8\,{\rm GHz}$.
Its negative spectral slope of $\alpha=-0.85$ is consistent with optically thin synchrotron emission.
The observed radio flux density of this source, $\sim10^2\,{\rm \mu Jy}$, is much higher than the free-free emission predicted in this work.
Therefore, if similarly bright radio emission were present in J1025+1402, the free-free signal could be completely overwhelmed.
However, J1025+1402 itself remains undetected with the \textit{VLA} \citep{Rodriguez2026}, suggesting that the presence and the strength of synchrotron radio emission may vary from object to object.

\cite{Rodriguez2026} also reported a $45\pm10\,{\rm \mu Jy}$ detection at a projected offset of $2^{\prime\prime}$ from J1025+1402, which has been discussed as a possible jet. 
This highlights the importance of spatially resolving the radio emission. 
High-spatial-resolution observations with \textit{ngVLA} may therefore be crucial for separating extended or offset jet radio emission from the compact free-free emission produced by the ionized gas surrounding the central engine.

The \textit{ngVLA} will provide still higher sensitivity over the overlapping frequency range and extend the coverage up to $\simeq116\,{\rm GHz}$ \citep{ngvla_2018}, making it particularly important for identifying the free-free spectral turnover.
At lower frequencies, \textit{SKA1-Low} will extend down to $\simeq50\,{\rm MHz}$ \citep{Braun2019}, providing complementary constraints on synchrotron emission and helping distinguish the free-free component from synchrotron emission.

\section{Conclusion}
In this Letter, we have investigated free-free radio emission as a new probe of ionized gas components in LRDs.
As illustrated in Figures\,\ref{fig:Lnuc_nuc} and \ref{fig:SED}, free-free self-absorption produces a characteristic spectral turnover.
Its critical frequency ($\nu_{\rm c}$) and luminosity ($L_{\nu_{\rm c}}$) can be used to constrain the electron column density, $N_{\rm e}$, and the characteristic size of the ionized region, $R$.
The spectral slope below $\nu_{\rm c}$ further probes the radial density profile.
Free-free emission thus provides a unique means of directly testing the presence of, and characterizing, dense ionized gas on pc and sub-parsec scales that cannot be spatially resolved even with future observatories, offering an independent test of the dense-gas and electron-scattering scenarios.

As demonstrated in Figures\,\ref{fig:model_detectability} and \ref{fig:SED} and the mock analysis in Figure\,\ref{fig:MCMC}, \textit{ALMA} and future \textit{ngVLA} can probe the range of ionized-gas properties.
Local LRD analogs provide particularly valuable targets, for which even \textit{ALMA} alone can constrain the free-free SEDs for cases with $R\gtrsim1\,{\rm pc}$ for $N_{\rm e}\simeq3\times 10^{23}\,{\rm cm^{-2}}$.
In the longer term, \textit{ngVLA}, planned for the 2030s, can extend such studies to higher redshifts.
These future observations would therefore provide an independent test of the physical nature of LRDs and the electron-scattering scenario, while probing the structure of dense ionized gas on spatial scales that are inaccessible through observations at other wavelengths.

\begin{acknowledgments}
We thank Taiki Kawamuro, Rohan Naidu, and Jialai Wang for fruitful discussions.
We thank Xiaojing Lin for sharing the spectrum of J1025+1402.
This work is based on observations associated with program ID 2561 and made with the NASA/ESA/CSA James Webb Space Telescope.
The data products presented herein were retrieved from the Dawn JWST Archive (DJA).
DJA is an initiative of the Cosmic Dawn Center (DAWN), which is funded by the Danish National Research Foundation under grant DNRF140.
Kavli IPMU is supported by World Premier International Research Center Initiative (WPI), MEXT, Japan.
T.S.T. is supported by Japan Society for the Promotion of Science (JSPS) KAKENHI Grant Number JP25KJ0750 and the Forefront Physics and Mathematics Program to Drive Transformation (FoPM), a World-leading Innovative Graduate Study (WINGS) Program at the University of Tokyo.
K.I. acknowledges support from the National Natural Science Foundation of China (12573015, W2532003), the Beijing Natural Science Foundation (IS25003), and the China Manned Space Program (CMS-CSST-2025-A09).
T.K. acknowledges support by KAKENHI (26KJ1232) through Japan Society for the Promotion of Science (JSPS).
OpenAI ChatGPT was used under author supervision to assist with language editing and refinement of the manuscript text. 
All scientific content, interpretations, references, and final wording were reviewed and approved by the authors, who take full responsibility for the content of the manuscript.

\end{acknowledgments}

\software{
astropy \citep{Astropy2013, Astropy2018, Astropy2022},
emcee \citep{emcee}
}
\bibliography{LRD}{}
\bibliographystyle{aasjournalv7}

\end{document}